\documentstyle[12pt]{article}
\input epsf
\begin{document}
\def\today{\space\number\day\ \ifcase\month\or
January\or February\or March\or April\or May\or June\or July\or
August\or September\or October\or November\or December\fi
\ \number\year}
\overfullrule=0pt  
\def\mynote#1{{[{\it NOTE: #1}]}}
\def\fEQN#1#2{$$\hbox{\it #1\hfil}\EQN{#2}$$}
\def\Acknowledgements{{\bigskip\leftline
{{\bf Acknowledgments}}\medskip}}

\begin{titlepage}

\begin{flushright}
DFTUZ/97-01\\  
 hep-ph/9702204\\

\end{flushright}

 \vspace{0cm}
 
\begin{center}
{\large\bf An extension of the standard model \\    
with a single coupling parameter }

 \vspace{0.4cm}

{\bf Mario Atance}\footnote{E-mail atance@posta.unizar.es},   
{\bf Jos\'e~Luis~Cort\'es}\footnote{E-mail cortes@posta.unizar.es}  
and {\bf Igor G. Irastorza}\footnote{E-mail igarcia@encomix.es}

 \vspace{0.2cm}

{\sl Departamento de F\'{\i}sica Te\'orica,\\ 
Universidad de Zaragoza,
50009 Zaragoza, Spain.}

 \vspace{0.2cm}

\centerline{ January 1997}

\end{center}
\vspace{0.2cm}
\begin{abstract}
We show that it is possible to find an extension of the
matter content of the standard model with a unification of
gauge and Yukawa couplings reproducing their known values. 
The perturbative renormalizability of the model with a single 
coupling and the requirement to accomodate the known properties 
of the standard model fix the masses and couplings of the 
additional particles. The implications on the parameters of 
the standard model are discussed.
  
\bigskip

\end{abstract}

\end{titlepage}

There are two main guidelines going beyond the standard model. The first 
one is based on consistency requirements which include as a crucial 
ingredient the incorporation of the gravitational interaction and it 
leads at the end to a candidate for the theory of everything (superstring 
theory, M-theory,...).
The problem with this approach is that, at least at present, the
understanding of the details how the standard model is contained in the
fundamental theory is not good enough to make testable predictions at
low energies. An alternative approach is to take the presence of a large 
number of free parameters in the standard model as the main motivation
to go beyond it considering possible extensions with a smaller number of
independent parameters. The main principle that has been used in the 
construction of extensions along this second line is an enlargement 
of the symmetries of the theory. The unification of gauge couplings in 
grand unified theories is not accompannied by a unification of the 
remaining couplings and the number of parameters is not substancially 
reduced. In the case of supersymmetric theories the situation is even 
worst because the unification of couplings is accompannied by the 
necessity to introduce additional free parameters describing the 
breaking of supersymmetry and one ends up with a theory with more free 
parameters than the standard model.

The unification of couplings through a symmetry is an example of
relations between renormalized couplings which are independent of 
the renormalization scale and compatible with the renormalization
group equations. The possibility to have a renormalizable theory 
(not necessarely based on a symmetry principle) with a reduced 
number of parameters (method of reduction of couplings) 
has been studied in recent years for different 
purposes\footnote{For a recent review with a list of references 
see~\cite{O}}. The program of reduction of couplings was
initiated in~\cite{OZ} by looking for massless
renormalizable theories in the power counting sense with a
single dimensionless coupling parameter. The same idea can
be applied in the case of effective field theories~\cite{eft}. 
Examples of phenomenological applications of the idea of 
reduction are the attempts to calculate quark masses within 
the framework of reduction of couplings in the standard model
~\cite{KSZ} and more recently in supersymmetric grand unified
theories~\cite{KMZ} as well as the unification of soft 
supersymmetry breaking parameters~\cite{JJ-KMZ2}.

In this paper we consider the principle of reduction of 
couplings as an alternative to the symmetry principle
in order to go beyond the standard model. The conjecture is that 
the theory beyond the standard model is such that its low energy
limit is described by a theory with a perturbative expansion in
terms of a single independent coupling. All the couplings of the 
theory can be expressed as a power expansion in a single coupling
with scale independent coefficients, valid at any scale above a 
mass $M$ identified as the reduction scale (reduction principle).
From the point of view of selection of solutions of the 
renormalization group equations the reduction principle corresponds
to a complete reduction of couplings, as refered in~\cite{KSZ}, but
only above a scale $M$ ; it should be distinguished from the more
general reduction of couplings~\cite{KSZ} which correspond to 
solutions of the renormalization group equations with a number 
of constants smaller than the number of couplings. The reduction
of couplings should also be distinguished from the relations
between couplings which reflect the independence of the infrared
renormalization group flow on the ultraviolet physics~\cite{SW}. 

We show that it is necessary to go beyond the matter content of 
the standard model in order to have a perturbatively renormalizable 
theory with a single coupling and, assuming that there are no additional 
gauge interactions, the reduction principle fixes a minimal extension of 
the standard model.

The perturbative reduction of couplings can be systematically 
studied order by order in perturbation theory. The starting point 
is the one-loop renormalization group equations for the gauge couplings,
Yukawa couplings and scalar self couplings of the standard model.
The structure of these equations allows to determine first the 
reduction of each gauge coupling independently of the remaining 
parameters. Then one can consider the reduction of the Yukawa couplings
from the corresponding renormalization group equations because they
depend only on the gauge couplings (already reduced at the previous 
step) and the Yukawa couplings. Finally one can consider the 
renormalization group equations and the reduction for the scalar sector. 
This separation in three steps of the reduction of couplings can be 
repeated order by order in a perturbative expansion of the reduction 
based on a loop expansion of the renormalization group equations.

The one-loop renormalization group equations for the three gauge
couplings $g_3$, $g_2$, $g_1$ of a theory with the local
$SU(3) \times SU(2) \times U(1)$ symmetry of the 
standard model are

\begin{equation}
16 \pi^2 \mu {dg_i \over d\mu} = b_i g_i^3
\label{rgegi}
\end{equation}

\noindent where the coefficients $b_i$ are fixed by the matter 
content of the model~\cite{CEL}

\begin{equation}
b_i = -{11 \over 3} S_1 ({\cal G}_{i}) + {4 \over 3} S_3 (F_{i})
+ {1 \over 6} S_3 (S_{i})
\label{bi}
\end{equation}

\noindent The constants $S_1 ({\cal G}_{i})$, $S_3 (F_{i})$
and $S_3 (S_{i})$ are defined, in terms of the structure constants 
$C_{i}^{abc}$ and the generators $T_{i}^{a}$ of the group 
${\cal G}_i$,  by the identities

\begin{eqnarray}
C_i^{acd} C_i^{bcd} &=&  S_1 ({\cal G}_{i}) \delta^{ab} \,, \\
Tr_{F_i}(T_i^{a} T_i^{b}) &=& S_3 (F_{i}) \delta^{ab} \,, \\
Tr_{S_i}(T_i^{a} T_i^{b}) &=& S_3 (S_{i}) \delta^{ab}
\,, \label{S}
\end{eqnarray}

\noindent and $F_i$($S_i$) is the representation of the fermions
(scalars).

Introducing $\alpha_i=g_{i}^2 /4\pi$, the reduction of
the three gauge couplings to a single coupling requires that

\begin{eqnarray}
\alpha_2 /\alpha_1 &=& b_1 /b_2 \, \nonumber \\
\alpha_3 /\alpha_1 &=& b_1 /b_3 \,.
\label{rgi}
\end{eqnarray}

\noindent This is only possible if the three coefficients $b_i$
have the same sign which is not the case in the standard model.
Then in order to have a perturbative reduction of couplings it
is neccessary to include additional matter fields such that the
sign of $b_2$ and $b_3$ becomes positive\footnote{A reduction of 
couplings based on a more general solution of the renormalization
group equations is not consistent with the value of the top quark 
mass~\cite{KSZ}}.

The minimal extension of the standard model that one can consider
in order to avoid the obstruction to a reduction of gauge couplings
is to add vector-like fermions in a non-trivial representation
of $SU(3)$ and $SU(2)$. In this way one has an extension whose 
mass is independent of the breaking of the $SU(2) \times U(1)$
symmetry and then can be made naturally larger than the 
Fermi scale (which is compatible with the absence of any signal
of this extension). Another consequence of the vector-like 
character of the extension is that no additional couplings are
introduced with the extension of the fermionic field content.

The perturbative reduction of gauge couplings leads to a constant
value for the two ratios of gauge couplings at scales larger than
the mass $M$ of the vector-like fermions. These ratios are given 
in terms of the first coefficients of the $\beta$-functions by 
(\ref{rgi}). At scales lower than $M$ the additional fermions 
decouple and the gauge couplings evolve according to the 
renormalization group equations of the standard model. The first 
nontrivial phenomenological test of the validity of the perturbative 
reduction of couplings is to find a representation for the
fermionic extension compatible with the known values of the gauge 
couplings.
\medskip
\begin{figure}[h!t]
  \centerline{\epsfxsize=12cm \epsfbox{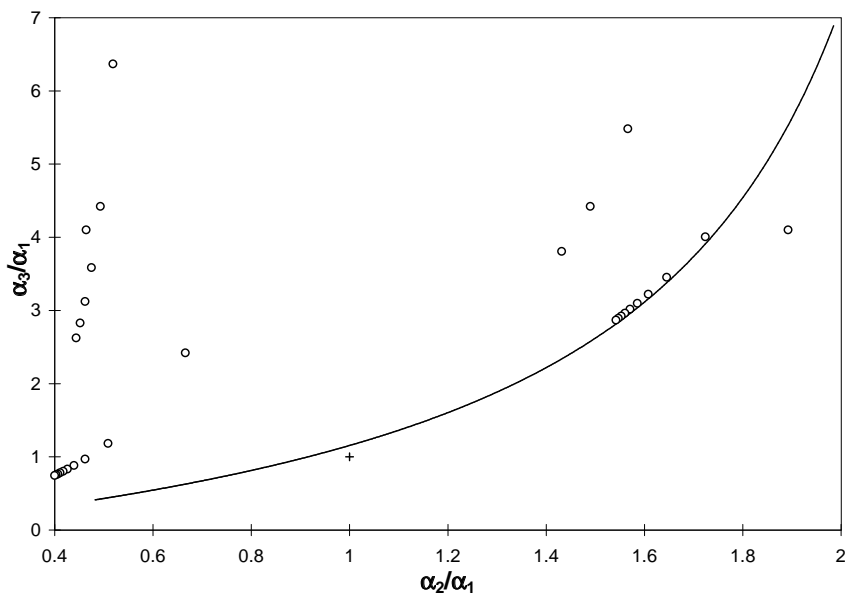}}
  \caption{{\small Ratios of gauge couplings in the standard 
  model as a function of the scale $\mu$ (curve starting on the rigth 
  at $\mu =M_{Z}$). Ratios of gauge couplings for a reduction of 
  couplings based on an extension of the standard model with
  additional fermions in a representation of
  $SU(3)\times SU(2)\times U(1)$ with a given multiplicity (points).
  $SU(5)$ unification (cross).}}
\end{figure}
\medskip
In fig.1 the values of the ratios $\alpha_2 /\alpha_1$
and $\alpha_3 /\alpha_1$ are plotted for (different multiplicities 
of) several representations of $SU(3) \times SU(2) \times U(1)$. We
have considered the lower dimensional representations of $SU(2)$ and 
$SU(3)$ up to the adjoint representation and integer values of
the $U(1)$-hypercharge $Y = 0,1,2$. Also represented in the figure is
the curve of values of these ratios in the standard model as a function 
of the renormalization scale $\mu$, for $M_Z \leq \mu \leq M_{Pl}$.
There are points on the curve, which means that the values of the gauge
couplings at low energies are compatible with the reduction of couplings,
and all of them correspond to different multiplicities $2\leq N\leq 10$
of a unique representation, $(8,3)$, of $SU(3)\times SU(2)$ with $Y=2$.
The different points on the curve lead to different values of the
scale $\mu =M$ of the fermionic extension increasing with the multiplicity.
For $N=3$ one has a vector-like representation for each generation and 
the mass of the additional fermions is $M\approx 70$ Tev. This result
for the unification of gauge couplings through a perturbative reduction
can be compared with the $SU(5)$-unification which corresponds to the
point $(1,1)$, close to the standard model curve at a much higher scale.

Once a phenomenologically valid reduction of gauge couplings has been 
identified one has to consider the renormalization group equations
for the Yukawa couplings. A vector-like fermionic extension does not
couple directly to the scalar field and then it does not introduce any 
modification at the one loop level on the renormalization group 
equations for the Yukawa couplings $y_f$ of the standard model~\cite{CEL}

\begin{eqnarray}
 16\pi^2 \mu {dy_{u_i} \over d\mu} &=& y_{u_i} \left[ \frac{3}{2} 
 y_{u_i}^2 - {3\over 2} y_{d_i}^2 
  + \sum_{j} \left( y_{l_j}^2 + 3 y_{u_j}^2 
 + 3 y_{d_j}^2 \right) \right.\nonumber \\
  & &\hspace{1cm}{}\left. - {17\over 20} g_1^2 - {9\over 4} g_2^2 
 - 8 g_3^2 \right] \,, \label{yu}
\end{eqnarray}

\begin{eqnarray}
 16\pi^2 \mu {dy_{d_i} \over d\mu} &=& y_{d_i} \left[ {3\over 2} 
 y_{d_i}^2 - {3\over 2} y_{u_i}^2 
  + \sum_{j} \left( y_{l_j}^2 + 3 y_{u_j}^2 
  + 3 y_{d_j}^2 \right) \right.\nonumber \\ 
  & &\hspace{1cm}{}\left. - {1\over 4} g_1^2 - {9\over 4} g_2^2 
  - 8 g_3^2 \right] \,, \label{yd}
\end{eqnarray}

\begin{eqnarray}
 16\pi^2 \mu {dy_{l_i} \over d\mu} &=&  y_{l_i} \left[ {3\over 2} 
 y_{l_i}^2 + \sum_{j} \left( y_{l_j}^2 + 3 y_{u_j}^2 
  + 3 y_{d_j}^2 \right) \right.\nonumber \\ 
  & &\hspace{1cm}{}\left.- {9\over 4} g_1^2 - {9\over 4} g_2^2 
  \right] \,. \label{yl}
\end{eqnarray}

\noindent where $u_i$, $d_i$, $l_i$ denote the quarks of charge
$2 /3$, $-1 /3$ and charged lepton of each generation
and the mixing between generations has been neglected.

Once more the reduction to a single coupling at the one loop
level requires the 
ratios $C_{u_i} = y_{u_i}^2 /g_1^2$, 
$C_{d_i} = y_{d_i}^2 /g_1^2$,
$C_{l_i} = y_{l_i}^2 /g_1^2$ to be independent of the 
renormalization scale. The consistency with the renormalization
group equations leads to the reduction equations

\begin{eqnarray}
 &C_{u_i}& \left[{3\over 2}C_{u_i}
 -{3\over 2}C_{d_i} 
 +\sum_{j}\left(C_{l_j} + 3 C_{u_j} 
 + 3 C_{d_j} \right) \right.\nonumber \\
  & & \left. -\left({17\over 20}+b_1 +
 {9\over 4}{b_1\over b_2}+8{b_1\over b_3}\right)\right] = 0
 \,, \label{Cui}
\end{eqnarray}

\begin{eqnarray}
 &C_{d_i}& \left[{3\over 2}C_{d_i}
 -{3\over 2}C_{u_i} 
 +\sum_{j}\left(C_{l_j} + 3 C_{u_j} 
 + 3 C_{d_j} \right) \right.\nonumber \\
  & & \left. -\left({1\over 4}+b_1 +
 {9\over 4}{b_1\over b_2}+8{b_1\over b_3}\right)\right] = 0
 \,, \label{Cdi}
\end{eqnarray}

\begin{eqnarray}
 &C_{l_i}& \left[{3\over 2}C_{l_i} 
 +\sum_{j}\left(C_{l_j} + 3 C_{u_j} 
 + 3 C_{d_j} \right) \right.\nonumber \\
  & & \left. -\left({9\over 4}+b_1 +
 {9\over 4}{b_1\over b_2} \right)\right] = 0
 \,. \label{Cli}
\end{eqnarray}

\noindent The only solution which is compatible with
the hierarchy of masses in the standard model corresponds
to take all the Yukawa couplings except $y_t$ equal to zero. 
The lowest order approximation for the mass,
$m_t^2 = y_{t}^2 \frac{v^2}{2}$ leads to $y_{t} (\mu = m_t) \simeq 1$
and, using the renormalization group equation of the top quark Yukawa 
coupling and the values of the gauge couplings at $\mu = M_{Z}$, one 
has that $16\pi^2 \mu \frac{dy_{t}^2}{d\mu} < 0$ over an
energy range including the scale $M$ identified in the reduction of 
gauge couplings. On the other hand the reduction to a single coupling 
implies that, for scales $\mu \geq M$, the squared top quark Yukawa 
coupling should grow with the renormalization scale $\mu$ as the 
gauge couplings do. 

Then in order to accomodate the value of the top quark mass one has to 
consider an extension of the standard model which changes the one loop 
renormalization group equation of $y_t$ or an extension where the 
relation between $y_t$ and $m_t$ of the standard model is modified.
In the first case one has to consider additional fermions with a 
direct coupling to the scalar doublet, i.e., additional chiral
fermions. The simplest possibility is to consider a new generation 
which gives an additional contribution of the required sign to
the $\beta$-function of $y_t$ through the renormalization of the 
scalar field. But in this case one has a new system of reduction 
equations for the Yukawa couplings of the new generation. 
The equations  (\ref{Cui}), (\ref{Cdi}) for the quarks of charge
$2 /3$ , $-1 /3$ of the forth generation will now include a
term proportional to $y_n^2$ from the contribution of the 
neutral lepton to the renormalization of the scalar field and the
reduction equations for the Yukawa couplings of the leptons will be

\begin{eqnarray}
 &C_{l_i}& \left[{3\over 2}C_{l_i} - {3\over 2}C_{n_i}
 +\sum_{j}\left(C_{l_j} + C_{n_j} + 3 C_{u_j} 
 + 3 C_{d_j} \right) \right.\nonumber \\
  & & \left. -\left({9\over 4}+b_1 +
 {9\over 4}{b_1\over b_2} \right)\right] = 0
 \,, \label{Cli2}
\end{eqnarray}

\begin{eqnarray}
 &C_{n_i}& \left[{3\over 2}C_{n_i} - {3\over 2}C_{l_i}
 +\sum_{j}\left(C_{l_j} + C_{n_j} + 3 C_{u_j} 
 + 3 C_{d_j} \right) \right.\nonumber \\
  & & \left. -\left({9\over 20}+b_1 +
 {9\over 4}{b_1\over b_2} \right)\right] = 0
 \,. \label{Cni}
\end{eqnarray}

\noindent The contribution due to the renormalization of the 
scalar field is the same for all the Yukawa couplings and, since
quarks and leptons are in the same representation of $SU(2)$, the
coefficients of the terms due to the scalar-fermion vertex
corrections are also the same. These two facts lead to the 
absence of any solution to the reduction equations with all the
Yukawa couplings of one generation different from zero as it is
required phenomenologically.

The other alternative to make compatible the value of the top 
quark mass with the reduction of couplings is to modify the scalar 
sector. The simplest possibility is to consider a two Higgs doublet 
model~\cite{2HDM}.
One still has the electroweak $\rho$ parameter equal to one in
the Born approximation and flavour-changing neutral currents
are avoided if all fermions with the same electric charge
couple to the same Higgs doublet~\cite{GW}. The renormalization 
group equations for the Yukawa couplings are in this case given by

\begin{eqnarray}
 16\pi^2 \mu {dy_{u_i} \over d\mu} &=& y_{u_i} \left[ \frac{3}{2} 
 y_{u_i}^2 + {1\over 2} y_{d_i}^2 
  + \sum_{j} \left( y_{n_j}^2 + 3 y_{u_j}^2 
  \right) \right.\nonumber \\
  & &\hspace{1cm}{}\left. - {17\over 20} g_1^2 - {9\over 4} g_2^2 
 - 8 g_3^2 \right] \,, \label{yu2}
\end{eqnarray}

\begin{eqnarray}
 16\pi^2 \mu {dy_{d_i} \over d\mu} &=& y_{d_i} \left[ {3\over 2} 
 y_{d_i}^2 + {1\over 2} y_{u_i}^2 
  + \sum_{j} \left( y_{l_j}^2 + 3 y_{d_j}^2 
  \right) \right.\nonumber \\ 
  & &\hspace{1cm}{}\left. - {1\over 4} g_1^2 - {9\over 4} g_2^2 
  - 8 g_3^2 \right] \,, \label{yd2}
\end{eqnarray}

\begin{eqnarray}
 16\pi^2 \mu {dy_{l_i} \over d\mu} &=&  y_{l_i} \left[ {3\over 2} 
 y_{l_i}^2 + {1\over 2} y_{n_i}^2 
  + \sum_{j} \left( y_{l_j}^2 + 3 y_{d_j}^2 
  \right) \right.\nonumber \\ 
  & &\hspace{1cm}{}\left.- {9\over 4} g_1^2 - {9\over 4} g_2^2 
  \right] \,, \label{yl2}
\end{eqnarray}

\begin{eqnarray}
 16\pi^2 \mu {dy_{n_i} \over d\mu} &=&  y_{n_i} \left[ {3\over 2} 
 y_{n_i}^2 + {1\over 2} y_{l_i}^2 
  + \sum_{j} \left( y_{n_j}^2 + 3 y_{u_j}^2 
  \right) \right.\nonumber \\ 
  & &\hspace{1cm}{}\left.- {9\over 20} g_1^2 - {9\over 4} g_2^2 
  \right] \,, \label{yn2}
\end{eqnarray}

\noindent where we have included a right-handed neutrino and a 
Yukawa coupling for the neutrino. The result for the chiral
fermion content of the minimal standard model is recovered by
considering three generations, $i = 1,2,3$ and making $y_{n_i} = 0$.
Once more the hierarchy of fermion masses requires to consider a 
solution with all Yukawa couplings except $y_t$ equal to zero. The
consistency of the reduction of couplings fixes the ratio
$C_t = y_t^2 /g_1^2$,

\begin{equation}
 C_t = \frac{17}{90} + \frac{2}{9}b_1 + \frac{1}{2}\frac{b_1}{b_2} +
 \frac{16}{9}\frac{b_1}{b_3} \,,
\end{equation}

\noindent for all the scales above the scale $M$ of the reduction 
of couplings which we have already identified at the level of 
gauge couplings. Using the values of the first $\beta$-function
coefficients $b_1$, $b_2$ and $b_3$ with one vector-like
representation $(8,3)_2$ for each generation and the 
renormalization group equation of the standard model for
$y_t$ for $\mu < M$ one finds $y_t (m_t) = 1.86$. In order
to reproduce the top quark mass it is neccesary to have

\begin{equation}
 y_t^2 (m_t) \frac{v_1^2}{v_1^2 + v_2^2} \simeq 1
 \,, 
\end{equation}

\noindent which in this case gives a condition for the v.e.v. of
the two doublets, $v_1$ and $v_2$, fixing the ratio
$\tan \beta = \frac{v_1}{v_2} \simeq .62$ . Then it is possible 
to accomodate the fermion masses (eleven massless fermions and
a massive top quark) in an extension of the standard model
with a single coupling parameter, based on the addition of a 
second Higgs doublet and a vector-like fermion representation 
$(8,3)_2$ of $SU(3) \times SU(2) \times U(1)$ for each generation
with a mass $M \simeq 70$ Tev. 

It is possible but not required to go beyond this minimal extension. 
For example one can consider additional generations which now due to 
the presence of a second Higgs doublet are compatible with a reduction 
of couplings. As an example of the predictive power of the reduction 
principle the masses of all the fermions of additional generations are 
fixed and turn out to be marginally compatible with present experiments 
due to the small values obtained for the mass of the neutrino of the 
additional generations. Another example corresponds to an extension of 
the standard model with only n generations of chiral fermions ($n > 6$ 
in order to have a reduction of gauge couplings) and two Higgs doublets. 
It can be excluded because the values of the gauge couplings at 
$\mu = M_{Z}$ are not consistent with the reduction of couplings. These 
examples make manifest the difficulty to find an extension of the 
standard model compatible with the reduction of couplings.

The last step in the renormalization group equation analysis leads us to
the one-loop equations for the scalar sector. Here one has 
seven dimensionless couplings and three parameters corresponding to the 
quadratic terms which appear in the more general potential 
with two Higgs doublets. The one loop $\beta$-functions for the 
couplings fix their reduction in terms of $g_1$ from the consistency
of the relations $\lambda_i = C_i g_1^2$, $i=1,...,7$ , with the 
renormalization group equations leading to a set of algebraic reduction 
equations for the constant coefficients $C_i$. Two combinations of the 
three parameters corresponding to the quadratic terms are fixed by the
required values of the two v.e.v. and as a consequence of the 
reduction to a single coupling it will be possible to determine the 
four masses and the mixing angle which characterize a two Higgs doublet
model in terms of one free parameter. A detailed discussion of
the predictions for the scalar sector toghether with a study of the 
possibility to go further in the reduction principle considering
a reduction of the three mass parameters in the potential to a single
one will be presented in a future work.

To summarize we have presented the reduction principle as an alternative 
to an enlargement of symmetries as a way to try to go beyond the standard 
model. The assumption behind the reduction of couplings is that the 
theory beyond the standard model is such that its low energy limit is as
independent on the details of the theory as possible. Since the details
of the theory appear at the level of the low energy limit through the 
values of the renormalized parameters it is the number of independent
couplings which have to be minimized. The arbitrariness of the extension
based on the symmetry principle which appears through the choice of 
symmetry group and representation of matter fields has an analog 
in the case of reduction of couplings in the choice of the matter 
field content of the extension once it is assumed that there are no 
additional gauge interactions. The mass scale of the extension, 
identified through the reduction to a single coupling, is many orders of
magnitud lower than the scale of grand unified theories and there is no
desert. As for the parameters of the standard model associated to the 
fermion masses (CKM~\cite{CKM} matrix and quark and lepton masses) they 
are in principle determined by the reduction. At the one loop level 
only one fermion has a non vanishing mass and there is no CKM matrix
but the reduction of couplings can be systematically extended order by
order in perturbation theory and the remaining masses and mixing of
generations could be a consequence of the reduction of couplings beyond 
one loop. This possibility as well as the extension of the reduction
of couplings to the effective theory including terms of dimension
greater than four in the Lagrangian which can also be done along the
same lines~\cite{eft} deserve further investigation.

\bigskip

\bigskip\leftline{{\bf Acknowledgments}}\medskip
\noindent We are gratefull to J.~L.~Alonso for discussions and
M.~Asorey for a reading of the manuscript. This work was partially 
supported by CICYT contract AEN 96-1670.  
 
\vfill
\eject

\newpage

\end{document}